\documentclass[a4paper,11pt]{article}
\usepackage{pos}

\usepackage{braket}
\usepackage{comment}

\title{Possible scenario of dynamical chiral symmetry breaking in the instanton liquid}
\author*[a]{Yamato Suda}
\author[a]{Daisuke Jido}

\affiliation[a]{Department of Physics, Tokyo Institute of Technology,\\
  2-12-1 Ookayama, Meguro, Tokyo 152-8551, Japan}

\emailAdd{suda.y.ad@m.titech.ac.jp}
\emailAdd{jido@th.phys.titech.ac.jp}

\abstract{
Based on simulations of the interacting instanton liquid model (IILM) 
with three-flavor quarks, we compute the free energy density 
of the QCD vacuum as a function of the quark condensate. 
We then evaluate the second derivative of 
the free energy density with respect to 
the quark condensate at the origin. 
This evaluation allows us to investigate whether 
chiral symmetry breaking in the IILM occurs 
in an anomaly-driven way. 
Such a breaking pattern of chiral symmetry 
has been proposed in a previous study 
to connect the QCD vacuum structure with 
meson properties, such as the mass of the sigma meson. 
We also perform the quenched simulations, 
in which no dynamical quarks interact with instantons.
Comparing these results with the full calculations 
provides a better understanding of 
the pattern of chiral symmetry breaking in the IILM.
We find that in the full IILM, 
chiral symmetry is dynamically broken in anomaly-driven way, 
whereas in the quenched IILM, 
it is broken through the ordinary mechanism. 
Based on these results, we suggest that chiral symmetry breaking 
in real QCD could also occur in an anomaly-driven way.
Consequently, in phenomena where chiral symmetry breaking plays 
a crucial role, the anomaly effect may also have significant influence.
}

\FullConference{10th International Conference on Quarks and Nuclear Physics (QNP2024)\\
8-12 July, 2024\\
Barcelona, Spain\\}

\begin{document}
\maketitle

\section{Introduction}
Spontaneous chiral symmetry breaking 
is a fundamental phenomenon in strong interaction,
playing a crucial role in hadron physics.
It governs key aspects such as the generation of hadron masses
and the emergence of massless Goldstone bosons.
Traditionally, chiral symmetry breaking is understood 
as follows: the attractive interaction mediated 
by gluon exchanges between quarks leads to the formation of 
a quark-antiquark condensate in the vacuum, 
and it breaks chiral symmetry spontaneously. 
This mechanism is widely described by the Nambu-Jona-Lasinio (NJL) model, 
an effective low-energy theory of quantum chromodynamics (QCD).

A recent study has revisited 
the relationship between the pattern of chiral symmetry breaking 
and the axial anomaly. One well-known example is the NJL model, 
which has been extensively studied ans is known to exhibit 
chiral symmetry breaking due to the attractive interaction between quarks, 
as described above. Quantitatively, this symmetry breaking occurs 
when the coupling $g_S$ of the inter-quark interaction 
exceeds a critical value $g_S^{\rm crit}$.
The previous study demonstrated that even when $g_S<g_S^{\rm crit}$, 
dynamical symmetry breaking can still occur if there is 
a sufficiently strong contribution from the axial anomaly, 
introduced via the Kobayashi--Maskawa--'t Hooft (KMT) term. 
In such cases, this anomaly-driven mechanism has been shown to 
influence physical quantities, 
such as causing the sigma meson mass to become 
lighter than $800~{\rm MeV}$~\cite{Kono:2021}. 

Following to the previous study,
we define patterns of chiral symmetry breaking, 
distinguishing between the conventional one known so far and
the one proposed in the previous study due to the contribution of the axial anomaly. 
We refer to the former breaking pattern as ``ordinary breaking'' 
and the latter as ``anomaly-driven breaking''. 
Instead of classifying them based on the magnitude of the coupling constants,
we adopt a more general approach. 
We define them by the sign of the second derivative of the vacuum energy density 
with respect to the quark condensate at the origin, 
where a negative (positive) value corresponds to ordinary (anomaly-driven) 
breaking~~\cite{Suda:2024}.

We examine whether the anomaly-driven breaking can occur in other 
model rather than those used in the previous study. 
We focus on the fact that the anomaly effect is understood 
as the instanton induced interaction. 
Since such effect might be naturally incorporated in the instanton liquid model, 
which is a phenomenological framework describing the QCD vacuum with instanton degree of freedom, 
we expect that chiral symmetry could be broken in the anomaly-driven way. 
The details of this study are presented in our paper~\cite{Suda:2024}. 

%In this presentation, we begin with a short introduction of the model in Sect.\ 2. In Sect.\ 3, 
%we explain the numerical calculation methods. In Sect.\ 4, we summarize 
%our calculation results, and in Sect.\ 5 we discuss their interpretation. 
%Finally, in Sect.\ 6, we give outlooks for wor work.  

\section{Partition function of the interacting instanton liquid model}

The partition function of the interacting instanton liquid model is given by \cite{Schafer:1996}
\begin{eqnarray}
  Z_{\rm IILM} = \sum_{N_+,N_-=0}^\infty \frac{1}{N_+! N_-!} 
  \int \prod_{i=1}^{N_+ + N_-} d\Omega_i f(\rho_i) e^{-S_{\rm int}} \prod_{f=1}^{N_f} \det (i\gamma^\mu D_\mu + m_f), 
\end{eqnarray}
where $N_+$ and $N_-$ are number of instantons and antiinstantons inside the four volume $V_4$, 
the measure of the path integral $d\Omega$ consists of size, color orientation and position of the $i$-th 
instantons $d\Omega_i= d\rho_i dU_i d^4z_i$, $f(\rho_i)$ represents 
the classical instanton amplitude, $S_{\rm int}$ is the instanton-instanton interaction and 
$D_\mu$ is covariant derivative for quark with current mass $m_f$ for each flavor $f$. 

\section{Numerical calculation method}

%We here shortly give the calculation methods in this work. 

\subsection{Vacuum energy density}
The vacuum energy density $\epsilon$ of the vacuum is identified as 
the free energy density $F$ at zero temperature ($T=0$), $\epsilon = F$. 
For simplicity, we will refer to it as ``free energy'' below.
From the standard thermodynamics relation, 
the free energy at $T=0$ is expressed as $F= - (1/V_4) \ln Z$ with the partition function of 
the system $Z$. It is clear that to obtain the vacuum energy density $\epsilon$ at $T=0$, 
it is sufficient to calculate the equivalent free energy $F$, and to calculate $F$, 
we only need to calculate $Z$. The partition function $Z$ can be calculated as follows. 
One writes the partition function as $Z(\alpha) = \int d \Omega \exp [-S_{\rm eff}(\alpha)]$ 
such that the desired partition function $Z$ is reproduced as $Z(\alpha=1)$. 
Here, an effective action $S_{\rm eff}(\alpha)$ has the form of $S_{\rm eff}(\alpha) = S_{\rm eff}^0 + \alpha S_1$ 
to interpolate between a known solvable action $S_{\rm eff}^0 \equiv S_{\rm eff}(0)$ 
and the full one $S_{\rm eff}(1)$. With this decomposition, 
one obtains the partition function $Z(\alpha=1)$ as follows:
\begin{eqnarray}
  \ln [Z(\alpha=1)] = \ln [Z(\alpha=0)] - \int_0^1 d\alpha \braket{0|S_1|0}_\alpha,
\end{eqnarray}
where an expectation value $\braket{0|O|0}_\alpha$ is defined by 
\begin{eqnarray}
  \braket{0|O|0}_\alpha = \frac{1}{Z(\alpha)} \int d\Omega O(\Omega) e^{-S_{\rm eff}(\alpha)}.
\end{eqnarray}
The explicit expression for the decomposition 
in our case is presented in Ref.~\cite{Suda:2024},
and it follows the one used in the previous study~\cite{Schafer:1996}.

\subsection{Quark condensate}
In our calculation using the IILM, the quark 
condensate for a single flavor $f$ with mass $m_f$ 
is evaluated as an expectation value of the traced 
quark propagator at the same point:
\begin{eqnarray}
  \braket{\bar{q}_fq_f} 
  = \sum_{A,\alpha} \braket{q^\dag_f (x)^A_\alpha q_f(x)^A_\alpha} 
  = - \lim_{y \to x} \frac{1}{Z} \int D\Omega {\rm Tr} [S(x,y;m_f)] e^{-S_{\rm int}} \det (\gamma^\mu D_\mu + m_f).
\end{eqnarray}
Here, we shortly write the measure in 
the path integral as $D\Omega$ given in the partition function.
$A=1,\dots,N_c$ and $\alpha=1,\cdots,4$ represent the color and the Dirac indices, respectively, 
and the trace operation is taken over for the both indices. 
The quark propagator $S(x,y;m)$ is approximated 
as a sum of contributions from the free and the zero-mode propagators
$S(x,y;m) \approx S_0(x,y) + S^{\rm ZM}(x,y;m)$~\cite{Schafer:1998}.
In our results, we only consider the zero-mode part of the propagator. 
Accordingly, the quark condensate is also considered only in terms of its zero-mode part. 
This is done to exclude the contribution of the free part, which diverges at the same point.

\subsection{Estimate of the curvature}
\label{sect.3.3}
We are interested in the curvature of the free energy density 
at the origin with respect to the quark condensate. 
Since the free energy and the quark condensate can be obtained 
as a function of the instanton density, 
combining these results, 
we have the free energy as a function of 
the quark condensate. 
Consequently, we can estimate 
the curvature at the origin of the free energy 
from these data. 
To obtain the curvature, we fit the data to a polynomial 
in the quark condensate of degree $K$: $F(x) = \sum_{j=0}^K C_j x^j$, 
where we write the quark condensate as $x$, 
and $C_j$ represents the coefficients resulting from the fitting. 
After fitting the data, we have the curvature 
at the origin of the free energy as the coefficient $C_2$.
In the actual fitting, we performed fits for cases 
where the polynomial degree $K$ are $2,3$ and $4$, 
and evaluated the systematic errors of the coefficients $C_j$.

\section{Numerical results}

We simulate the partition function 
with fixed number of instantons, $N=N_+ + N_-,$ using 
standard Monte-Carlo techniques. 
By varying the four-volume 
of the simulation while keeping $N$ fixed, 
we control the instanton density.
We perform two types of simulations.
The first is a full simulation,
which includes both inter-instanton and instanton-quark 
interactions. The second is a quenched simulation,
where the partition function 
omits the quark determinant, thereby excluding
the instanton-quark interaction from the system. 
For each simulation type, we perform the simulations 
with different quark masses $m_f$. 
In the full calculations, we use the quark masses of
$m_f = 37.0, 54.0$ and $70.0~{\rm MeV}$
in the flavor ${\rm SU(3)}$ limit. 
In contrast, for the quenched calculations, 
we use smaller quark masses of $m_f=2.8, 14$ and $28~{\rm MeV}$. 

\begin{figure}
  \centering
  \begin{minipage}{0.49\columnwidth}
    \includegraphics[width=60mm]{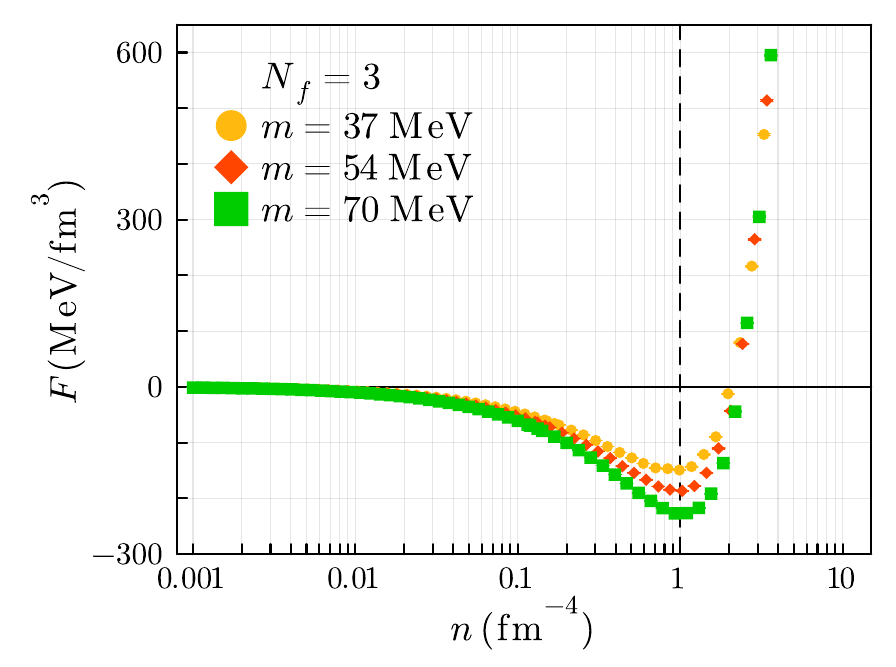}
    \caption{
      \label{fig:F_vs_n_unquench}
      Free energy as a function of 
      the instanton density.
      This figure is based on the work~\cite{Suda:2024}.
    }
  \end{minipage}
  \begin{minipage}{0.49\columnwidth}
    \includegraphics[width=60mm]{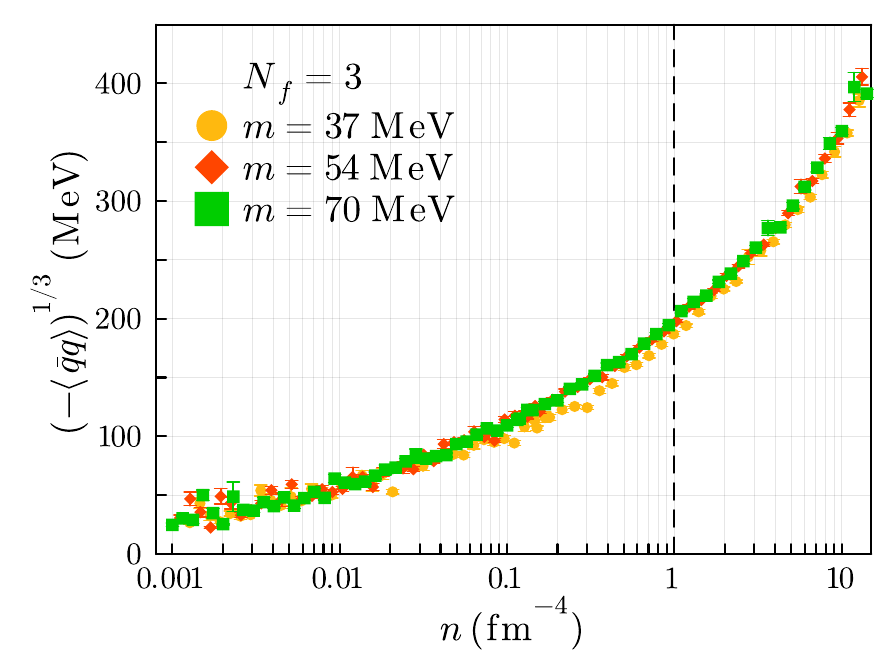}
    \caption{
      \label{fig:qq_vs_n_unquench}
      Quark condensate as a function 
      of the instanton density. 
      This figure is based on the work~\cite{Suda:2024}.
    }
  \end{minipage}
\end{figure}
\begin{figure}
  \centering
  \begin{minipage}{0.49\columnwidth}
    \includegraphics[width=65mm]{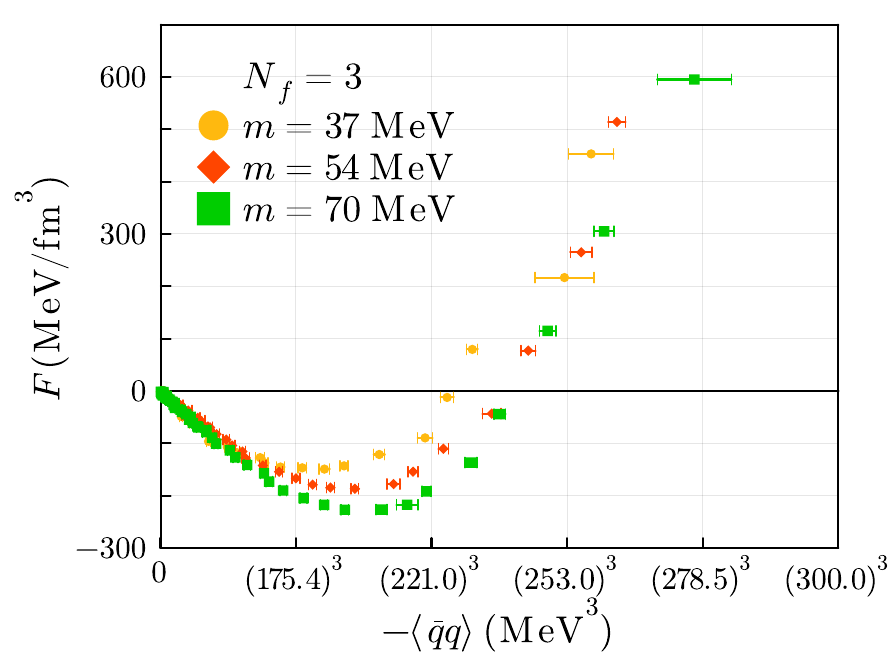}
    \caption{
      \label{fig:F_vs_qq_unquench}
      Free energy versus quark condensate.
      This figure is based on the work~\cite{Suda:2024}.
    }
  \end{minipage}
  \begin{minipage}{0.49\columnwidth}
    \includegraphics[width=55mm]{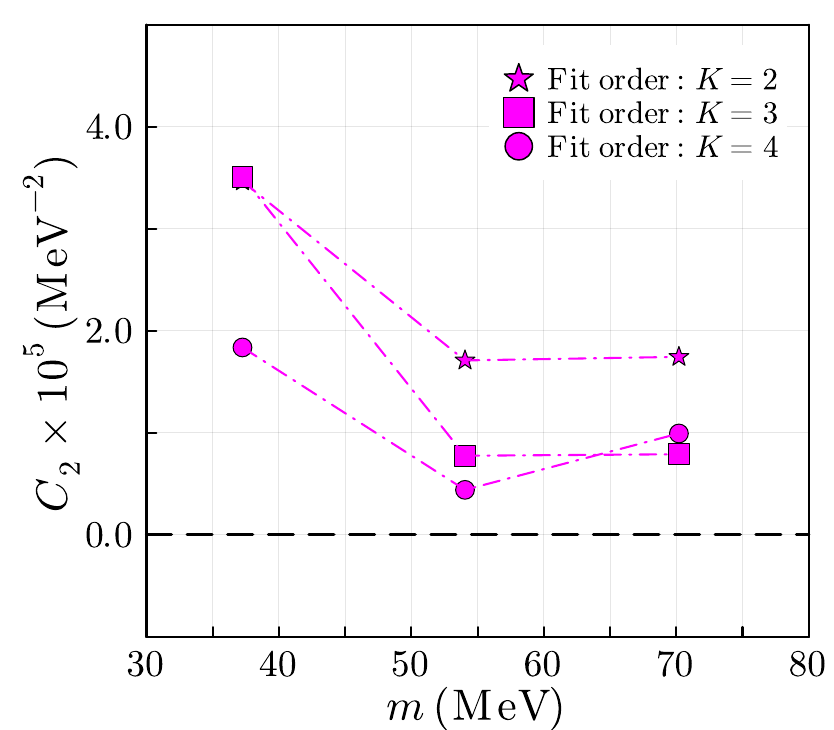}
    \caption{
      \label{fig:C2_vs_mq_unquench}
      Coefficient $C_2$ 
      for different current quark masses.
      This figure is based on the work~\cite{Suda:2024}.
    }
  \end{minipage}
\end{figure}

\subsection{Full calculation}
Figure \ref{fig:F_vs_n_unquench} shows 
the free energy as a function of 
the instanton density in the full calculation
for three different quark masses. At lower instanton density,
instantons feel attractive interaction and 
so the free energy decreases with increasing the density. 
While for higher density, the free energy rapidly grows. 
That shows repulsive interaction between instantons. 
Figure \ref{fig:qq_vs_n_unquench} shows 
the quark condensate in magnitude 
as a function of the instanton density
in the full calculations. The magnitude 
of the quark condensate increases 
with growing the instanton density, 
which reproduces well the previous work~\cite{Schafer:1996}.
Figure \ref{fig:F_vs_qq_unquench} shows 
the free energy versus the quark condensate 
in the full calculation. We find 
that chiral symmetry broken at the vacuum of the system
where minimizing the free energy.
Figure \ref{fig:C2_vs_mq_unquench} shows 
the curvature versus the current quark masses
in the full calculations. The curvature is evaluated 
as in the manner that we explained in Sect.~\ref{sect.3.3}. 
One can see that the value of $C_2$ is positive 
for the range we considered and for degree $k$ of polynomial 
in our full calculation of the IILM. This suggests that 
the IILM shows the anomaly-driven breaking of chiral symmetry.

\subsection{Quenched calculation}
\begin{figure}
  \centering
  \begin{minipage}{0.49\columnwidth}
    \includegraphics[width=65mm]{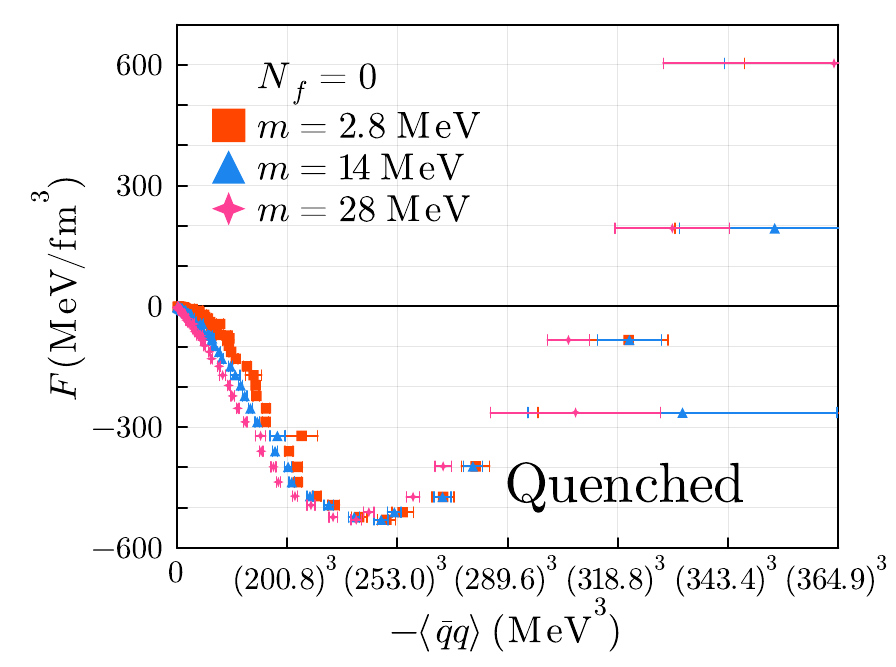}
    \caption{
      \label{fig:F_vs_qq_quench}
      Free energy versus quark condensate.
      This figure is based on the work~\cite{Suda:2024}.
    }
  \end{minipage}
  \begin{minipage}{0.49\columnwidth}
    \includegraphics[width=55mm]{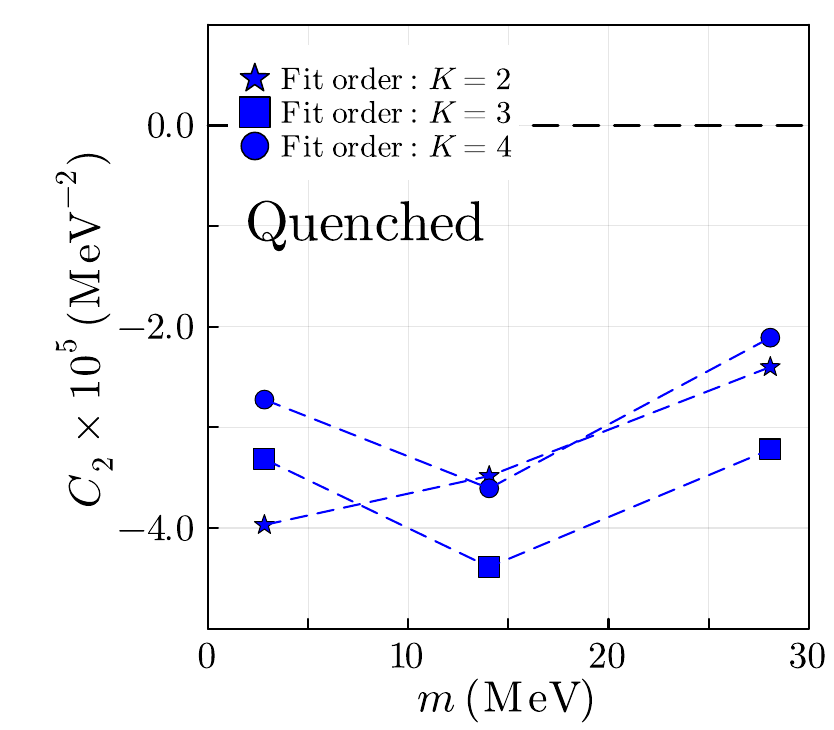}
    \caption{
      \label{fig:C2_vs_mq_quench}
      Coefficient $C_2$ 
      for different current quark masses.
      This figure is based on the work~\cite{Suda:2024}.
    }
  \end{minipage}
\end{figure}

We perform the same analysis for the quenched calculations.
Figure~\ref{fig:F_vs_qq_quench} shows 
the quenched calculation results 
for the free energy versus the quark condensate.
From this, we again observe that 
chiral symmetry is broken at the vacuum of the system. 
We can see that the decrease in the free energy 
as a function of the quark condensate is more gradual 
compared to the case of the full calculation results.
Figure~\ref{fig:C2_vs_mq_quench} shows 
the curvature $C_2$ versus the current quark mass 
in the quenched calculations.
We find the negative value of the curvature 
for a wide range of quark mass, and then we conclude 
that the quenched IILM shows the ordinary symmetry breaking pattern. 

\section{Discussion}
The realization of the anomaly-driven breaking in the 
full IILM can be understood as follows. 
We have defined the anomaly-driven breaking of chiral symmetry 
as a situation where chiral symmetry is broken at the vacuum 
and the curvature of the vacuum energy density is positive at the origin.
As investigated in previous work~\cite{Kono:2021}, 
such a situation occurs in the three flavor NJL model when 
the attractive interaction between quarks is not strong enough, 
but the KMT term has sufficient strength. 
The KMT term, which induces the anomaly effect, 
includes a six-quark interaction, 
and this can be understood as part of the 't~Hooft vertex for three flavor~\cite{Shifman:1980}. 
This 't~Hooft vertex is incorporated into the quark determinant 
in the IILM. Therefore, the full calculation 
that takes the quark determinant into account should include 
the 't~Hooft vertex. Consequently, it can be naturally 
interpreted that the IILM exhibits anomaly-driven chiral symmetry breaking. 

The results of the quenched calculations can be interpreted as supporting this 
interpretation. In the quenched calculations, the quark determinant 
is set to $1$, which suppresses the effects of the instanton-quark interaction. 
As a result, the interaction corresponding to the six-quark KMT term is 
no longer included, and thus, anomaly-driven chiral symmetry breaking does not occur.
This can be interpreted as manifesting in the negative curvature $C_2<0$ 
for the quenched calculation.

\section{Outlook}
We are interested in two directions for the further development of our work.
The first is the investigation of chiral symmetry breaking patterns 
in intermediate flavor regimes, such as $N_f=2$. Six-quark interactions like 
the KMT term appear when considering three quark flavors, while they do not 
manifest in cases with $N_f=0,1,$ or $2$. So, we are interested in understanding what
chiral symmetry breaking patterns realize in such intermediate flavor scenarios.
The second is to examine whether the consequence of the anomaly-driven breaking 
can also be reproduced in the IILM. 
The previous work has shown that the sigma meson as the chiral partner of the pion 
becomes lighter than about $800~{\rm MeV}$. 
This is an important exploration for connecting the discussion based on the sign 
of the curvature of the free energy with actual physical quantities. 

\section{Acknowledgements}
%We thank Masayasu Harada and Kotaro Murakami for useful discussions. 
This work by Y.S. is supported by JST SPRING, Grant No.\ JPMJSP2106. 
The work of D.J. was supported in 
part by Grants-in-Aid for Sicentific Reserach from JSPS (Grants No.\ JP21K03530, No.\ JP22H04917, 
and No.\ JP23K03427).


\begin{thebibliography}{99}

\bibitem{Kono:2021}
  S.~Kono, D.~Jido, Y.~Kuroda, and M.~Harada,
  ``The role of U$_{A}$(1) breaking term in dynamical 
  chiral symmetry breaking of chiral effective theories,''
  PTEP {\bf 2021}, no. 9, 093D02 (2021).

\bibitem{Suda:2024}
  Y.~Suda and D.~Jido, 
  ``Possible scenario of dynamical chiral symmetry breaking in the interacting instanton liquid model,''
  Phys.\ Rev.\ D {\bf 110} (2024) 014037.

\bibitem{Schafer:1996}
  T.~Sch\'afer and E.~V.~Shuryak, 
  ``Interacting instanton liquid model in QCD at zero and finite temperature,''
  Phys.\ Rev.\ D {\bf 53} (1996) 6522.

\bibitem{Schafer:1998}
  T.~Sch\'afer and E.~V.~Shuryak, 
  ``Instantons in QCD,''
  Rev.\ Mod.\ Phys.\ {\bf 70} (1998) 323.

\bibitem{Shifman:1980}
  M.A.~Shifman, A.I.~Vainshtein, V.I.~Zakharov,
  ``Instanton density in a theory with massless quarks,''
  Nucl.\ Phys.\ B {\bf 163} (1980) 46.

\end{thebibliography}
\end{document}